# BP$_5$ Monolayer with Multiferroicity and Negative Poisson's Ratio: A Prediction by Global Optimization Method

Haidi Wang,[†] Xingxing Li,[†,‡] Jiuyu Sun,[†] Zhao Liu[†] and Jinlong Yang[†,‡,*]

[†]Hefei National Laboratory for Physical Sciences at the Microscale, University of Science and Technology of China, Hefei, Anhui 230026, China

[‡]Synergetic Innovation Center of Quantum Information & Quantum Physics, University of Science and Technology of China, Hefei, Anhui 230026, China

Based on variable components global optimization algorithm, we predict a stable two-dimensional (2D) phase of boron phosphide with 1:5 stoichiometry, i.e. boron pentaphosphide (BP$_5$) monolayer, which has a lower formation energy than that of the commonly believed graphitic phase (g-BP). BP$_5$ monolayer is a multiferroic material with coupled ferroelasticity and ferroelectricity. The predicted reversible strain is up to 41.41%, which is the largest one among all reported ferroelastic materials. Due to the non-centrosymmetric structure and electronegativity differences between boron and phosphorus atoms, an in-plane spontaneous polarization of 3.26×10$^{-10}$ C/m occurs in BP$_5$. Moreover, the recently hunted negative Poisson's ratio property, is also observed in BP$_5$. As an indirect semiconductor with a band gap of 1.34 eV, BP$_5$ displays outstanding optical and electronic properties, for instance strongly anisotropic visible-light absorption and high carrier mobility. The rich and extraordinary properties of BP$_5$ make it a potential nanomaterial for designing electromechanical or optoelectronic devices, such as nonvolatile memory with conveniently readable/writeable capability. Finally, we demonstrate that AlN (010) surface could be a suitable substrate for epitaxy growth of BP$_5$ monolayer.

In the past several years, two-dimensional (2D) materials with few layer atom thickness have been paid great attention due to their potential applications in electronics, optoelectronics, and energy conversion.[1] For instance, monolayer BN,[2] MoS$_2$[3–5] and Group IV monochalcogenides[6] are recently reported to possess large piezoelectricity[7] which would allow efficient mechanical-to-electrical energy conversion. Black phosphorene has been predicted to exhibit ferroelasticity,[8] which has potential application for designing nonvolatile memory devices. In addition, δ-P is predicted to be an auxetic material with a highly negative Poisson's ratio.[9] Auxetic materials have also attracted intense research interest recently,[10–12] since these materials possess some novel properties such as enhanced toughness and enhanced sound or vibration absorption.[13] Although 2D materials mentioned above usually possess outstanding properties, it's temporarily hard to adapt multifunctional application. Therefore, in order to integrate multi-functionality in 2D semiconductor devices, 2D materials that hold simultaneously two or more primary properties or functions are highly desirable.

As a new class of 2D materials, boron-phosphorus binary compound semiconductors have gained great attention in recent years. In experimental aspect, several thin film growth methods for synthesizing B-P binary compounds have been proposed, including chemical vapor deposition (CVD)[14–16] close-spaced vapor transport (CVT),[17] flux growth,[18] high pressure flux method,[19] and epitaxy growth.[20] However, no experiment gave the clear structure information of layered B-P binary compounds. In theoretical aspect, all the studies[21–23] were prone to use the graphitic structure of boron-phosphide (g-BP) as the most stable phase, which may be wrong in reality. Therefore, a thorough investigation of the crystal structure of boron-phosphide will not only provide insight into the structure information of experimentally synthesized B-P compounds, but also may lead to the discovery of new functional 2D materials.

In this work, we systematically predict the lowest-energy structures of 2D B-P compounds by using first principles simulation and variable components global optimization based on Cuttlefish algorithm.[24] A new phase with high concentration of phosphorus element, namely boron pentaphosphide (BP$_5$), is obtained, which has a lower formation energy than that of commonly believed g-BP phase.[21,22,25–27] The calculations present that BP$_5$ monolayer is an unprecedented 2D material with five primary properties simultaneously, i.e. ferroelasticity, ferroelectricity, negative Poisson's ratio, strongly anisotropic visible-light absorption and semiconducting property with high carrier mobility, which make BP$_5$ monolayer a candidate material for designing nonvolatile memory devices. Finally, the AlN (010) surface is predicted to be a potential substrate for epitaxy growth of BP$_5$ monolayer.

## Results

**Structure and stability.** We carry out variable components global optimization searching for different stoichiometric ratio $B_mP_n$ {m≤5 and n≤6} compounds. Then the phase stabilities of different B-P system are evaluated firstly by judging the formation energy of $B_mP_n$. If the formation energy of a compound is negative, this compound is considered to be stable with respect to decomposition into the elements. Clearly, the formation energy of $B_mP_n$ per atom is written as:[28]

$$\Delta E(B_mP_n) = \frac{E(B_mP_n) - m\mu_B - n\mu_P}{m+n}$$

where $\Delta E(B_mP_n)$ is the formation energy per atom; $E(B_mP_n)$ is the total energy of $B_mP_n$ calculated by DFT. The chemical potential of B and P ($\mu_B$ and $\mu_P$) are taken from the cohesive energy of α-B[29] and black phosphorene,[30,31] respectively. To account for all thermally stable structures, we construct the 'convex hull' or 'global stability line' of all considered binary phases.[32] As shown in **Fig.** 1a, the formation energy $\Delta E(B_m P_n)$ is plotted as a function of molar fraction of boron element, and all points on the convex hull (solid line) are stable against all decomposition reactions. It can be found that most of the B-P compounds have positive formation energy (above the dashed line), however, the g-BP, BP$_5$ and two metastable phases of BP$_5$ (See Supplementary Information **Fig. S2**) have the negative ones. Here, we mainly focus on newly predicted BP$_5$ phase (**Fig. 1b**). Similar to black phosphorene, BP$_5$ has a puckered sheet of linked atoms with a puckered thickness d=3.840Å. The space group is Pm2m. The optimized lattice parameters are *a*=4.635Å and *b*=3.278Å. From top view, it can be found that BP$_5$ is composed of six-membered rings with boron and phosphorus atoms. The side view shows that BP$_5$ has a sandwich structure with a B-P single layer encapsulated by phosphorus atoms. It is worth noting that the formation energy of BP$_5$ is about 0.013 eV/atom lower than that of g-BP, which implies that BP$_5$ is thermally more stable than the commonly believed g-BP.

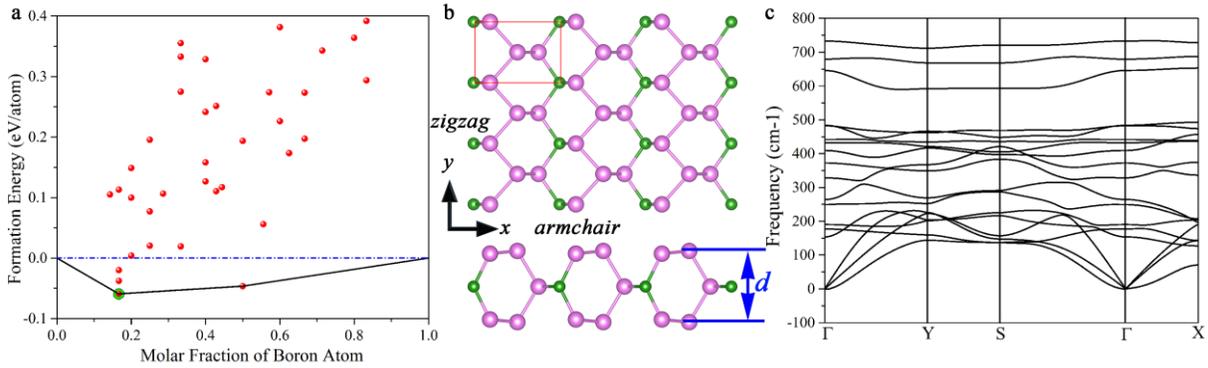

**Figure 1 | Geometric structure and stability of BP$_5$ monolayer. (a)** The formation energy per atom as a function of molar fraction of boron atom for different B-P phases where solid line denotes the convex hull and BP$_5$ is marked by green circle. **(b)** The top and side view of BP$_5$, where purple and green spheres denote P and B atoms. *d* stands for the puckered thickness of monolayer BP$_5$ and red box indicates the unit cell. **(c)** The phonon band structure of BP$_5$. *Γ* (0.0, 0.0, 0.0), *X* (0.5, 0.0, 0.0), *Y* (0.0, 0.5, 0.0) and *S* (0.5, 0.5, 0.0) refer to special points in the first Brillouin zone.

As discussed above, a new phase BP$_5$ is proposed by global search algorithm. To further check the dynamical stability, the phonon band structure is calculated. The absence of imaginary mode in the whole 2D reciprocal space indicates that this monolayer is dynamically stable **Fig. 1c**. Then ab initio molecular dynamics (AIMD) simulations are carried out with a 5×7 supercell to verify the thermal stability of BP$_5$ under finite temperature. After heating at 300K for 5ps (**Fig. S4**) with a time step of 1fs, the structure does not suffer significant distortion or transformation, so does it at 400K. In addition, the calculated elastic constants meet the necessary mechanical equilibrium conditions[33] for mechanical stability: $C_{11}C_{22}-C_{12}^2>0$ and $C_{11}$, $C_{22}$, $C_{66}>0$ (See Supplementary Information Table S1). Therefore, all these analysis above demonstrates that BP$_5$ is not only dynamically stable but also thermally and mechanically stable.

**Ferroelasticity.** Since BP$_5$ is predicted as a potential stable phase of low dimensional B-P binary compound, a further study of its intrinsic properties is desired. Firstly, we pay our attention to the ferroelastic switching of BP$_5$. A ferroelastic material is defined by the existence of two more equally stable orientation variants, which can be switched from one variant to another without diffusion by the application of external stress.[34,35] The two orientation variants of BP$_5$ are shown in **Fig. 2a**. As for the initial variant (I), the zigzag rows lie in y direction, and the lattice constants *b* along zigzag direction is less than that along armchair direction (*a*). If an external tensile stress is applied along the zigzag direction, it may transform into the variant III that will be energetically more favorable, where *a'*=*b*, *b'*=*a* and zigzag rows are switched to the x direction, same as the initial variant I upon 90 ° rotation. The inverse rotation can also be realized if the external tensile stress is applied along the zigzag direction for variant III. To estimate the transformation process between variant I and III, the pathway for the mechanical conversion **Fig. 2b** is computed by solid-sate nudged elastic band (ss-NEB) method.[36] The pathway from initial variant I to the final variant III is connected by a transition state II. The calculated energy barrier is about 0.32 eV/atom (the corresponding stress is 15.8 GPa, see **Fig. S5**), which is comparable with that value of BP (0.2 eV/atom).[8] Apart from energy barrier, the reversible ferroelastic strain [defined as (*a*/*b*-1)×100% ] is also a judgment criterion for performance of ferroelastic material, because higher reversible strain will exhibit stronger signal of switching. Compared with previously proposed ferroelastic materials,[8] $BP_5$ displays a reversible ferroelastic strain up to 41.4%, which is the highest value known to date.

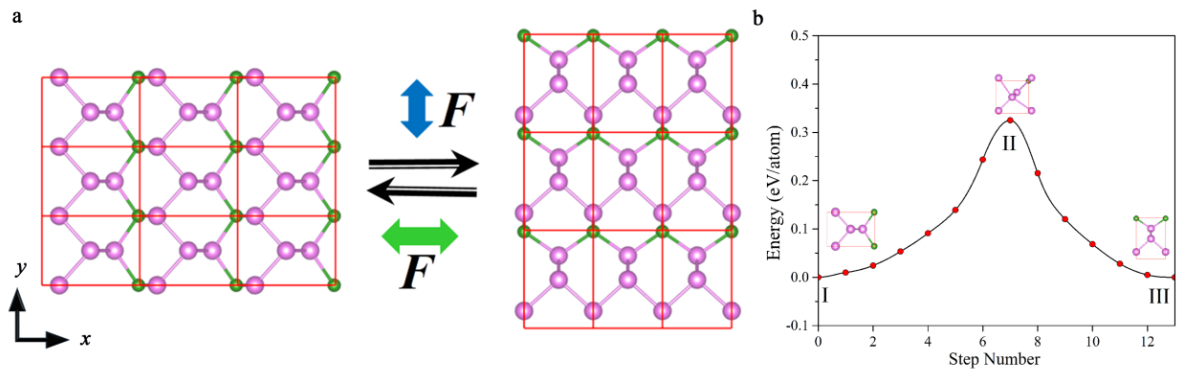

**Figure 2 | Ferroelastic properties of BP$_5$ monolayer. (a)** The ferroelastic switching for BP$_5$. The green and blue arrow denote the direction of tensile stress **(b)** The pathway of ferroelastic switching for BP$_5$ as a function of step number in solid-sate nudged elastic band (ss-NEB).

**Ferroelectricity.** Then we turn to discuss the ferroelectricity of $BP_5$. Unlike black phosphorene, $BP_5$ monolayer is non-centrosymmetric and the electronegativity of phosphorus atom (2.19) is larger than that of boron atom (2.04). As a result, a spontaneous polarization occurs in $BP_5$ monolayer with the direction of polarization along the +x direction. The estimated intensity of polarization is 3.26×10$^{-10}$ C/m, which is larger than that of SnS (2.47×10$^{-10}$ C/m) and SnSe (1.87×10$^{-10}$ C/m).[8] When an external field is applied along the +y direction, the boron atoms with positive charge will move upward and phosphorus atoms with negative charge will move downward. Finally, the variant III will be obtained with the direction of polarization along the +y direction. Because of the symmetry, the reaction path from III to I should be identical, and the ferroelectric switching follows the same path as the ferroelastic switching shown in **Fig. 2b**. Therefore, the activation barrier for ferroelectric switching and for ferroelastic switching is equivalent.

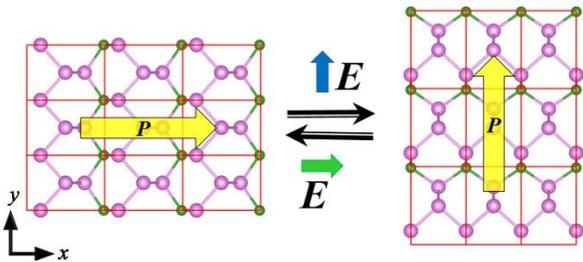

**Figure 3 | Ferroelectric properties of $BP_5$ monolayer.** Ferroelectric switching for $BP_5$ monolayer. The green and blue arrow denote the direction of external electric fields, while the yellow arrow display the spontaneous polarization direction.

**Negative Poisson's ratio.** Our previous work[9] has demonstrated that hinge-like structures, such as SnSe,[37] black phosphorene[11] and δ-P have the auxetic effect[10,11,38,39]. Due to the similar geometric structure, it is expected that auxetic effect may also exist in $BP_5$. **Fig. 4** shows that $BP_5$ indeed expands along out-of-plane direction when an external tensile stress is applied in zigzag direction. This indicates that $BP_5$ also holds a fascinating auxetic effect. Further calculation estimates that the negative linear Poisson's ratio is ν=-0.037 along the out-of-plane direction, which is a little higher than that of black phosphorene (ν=-0.027). Recently, an experimental work[40] has successfully demonstrated the negative Poisson's ration of black phosphorene, thus it is expected that auxetic effect of $BP_5$ could be validated by experiment in future.

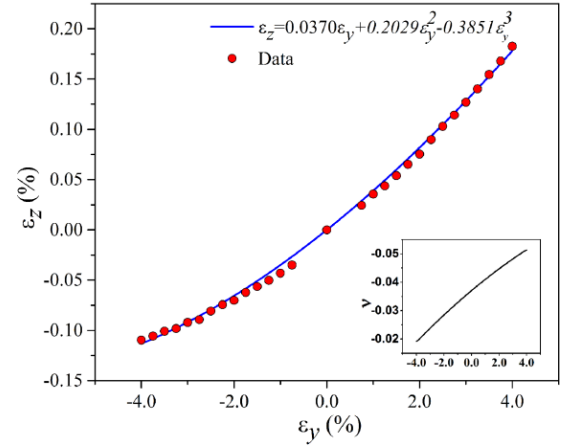

**Figure 4 | Negative Poisson's ratio of $BP_5$ monolayer.** The strain of out-of-plane direction versus strain along the y direction. Data is fitted to function $\varepsilon_z = 0.0370\varepsilon_y + 0.2029\varepsilon_y^2 - 0.3581\varepsilon_y^3$. The linear Poisson's ratio in the out-of-plane direction equals -0.037; Inset displays the out-of-plane Poisson's ratio as a function of strain obtained by $\nu = -\frac{d\varepsilon_z}{d\varepsilon_y}$.

**Electronic structure and optical properties.** The ferroelasticity and ferroelectricity of $BP_5$ discussed above will have more profound impact if they are also coupled to electronic and optical properties. The band structure and density of states (DOS) calculated under GGA-PBE level (See Supplementary Information S6) indicate $BP_5$ to be a semiconductor with an indirect band gap of 0.328 eV. The valence band maximum (VBM) is at Γ point and the conduction band minimum (CBM) is located between Y and Γ points. The decomposed DOS demonstrates that the VBM is mainly contributed by $p_z$-orbital of phosphorus atoms, while the CBM is a hybrid-state of phosphorus and boron atoms. Since the GGA-PBE functional significantly underestimates band gaps, the quasiparticle effect is then involved based on many-body perturbation theory within the $G_0W_0$ approximation to obtain the accurate band gap.[41,42] As shown in **Fig. 5a**, the indirect band gap of $BP_5$ is increased to 1.34 eV and the corresponding direct band gap is 2.53 eV. At the same time, the photoabsorption spectrum is also calculated by solving the Coulomb Bethe-Salpeter equation[43,44] based on the quasiparticle energies and the screened Coulomb interactions obtained from the GW calculations. **Fig. 5b** displays the calculated imaginary dielectric functions including the excitonic effect. $BP_5$ monolayer presents a highly anisotropic photoabsorption with two giant low-energy excitonic peaks: one at 1.60 eV for *yy* component and the other at 1.85 ev for the *xx* component of the dielectric tensor. Noting their corresponding direct quasiparticle transition energy is 2.54 eV and 2.62 eV respectively, we evaluate large exciton binding energies of 0.94 eV and 0.77 eV. These excitonic behaviors and strongly anisotropic absorption are similar to other 2D materials.[45]

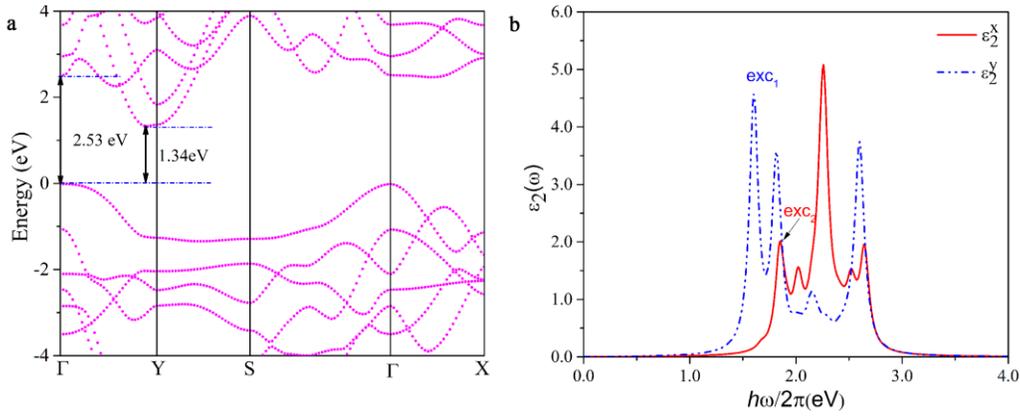

**Figure 5 | Electronic and optical properties of BP$_5$ monolayer. (a)** Band structure of BP$_5$ obtained within the G$_0$W$_0$ approximation **(b)** Imaginary dielectric function along the x (red line) and y (blue dotted line) direction.

**Table 1.** The calculated effective masses (m$^*$), deformation potential constants (E$_l$) and carrier mobilities (μ$_{2D}$) along the zigzag and armchair direction at T=298K with the GGA-PBE level of theory.

| Carrier | m* (m$_0$) | | E$_l$ (eV) | | μ$_{2D}$ (10$^3$cm$^2$/V s) | |
|---|---|---|---|---|---|---|
| | zigzag | armchair | zigzag | armchair | zigzag | armchair |
| electron | 0.346 | 0.428 | 1.284 | -5.032 | 7.069 | 0.608 |
| hole | 3.586 | 0.273 | -0.917 | -3.155 | 0.520 | 0.942 |

The inherent orthorhombic waved structure of BP$_5$ also results in the orientation-dependent effective carrier masses. As summarized in **Table 1**, hole has a large effective mass of 3.586$m_0$ and electron has a small value of 0.346$m_0$ along the zigzag direction. However the effective masses of electron and hole along the armchair direction have relatively small values, which are comparable to the values of phosphorene (0.17 $m_0$ and 0.15 $m_0$).[46,47] Carrier mobility is known to be a crucial factor for electronic and optoelectronic applications because high mobility can prevent electrons and holes from recombination[48] and have promising applications to future high-performance electronic devices.[49] Therefore, we calculate the carrier mobility based on the deformation potential (DP) theory as proposed by Bardeen and Shockley (See Supplementary Information).[50] The calculated carrier mobility is orientation-dependent as one would expect from the anisotropic nature of the calculated effective masses and mechanical properties. For example, the mobility of electrons along the zigzag direction is higher than that of armchair one. The underlying reason is that the deformation potential for electrons along the zigzag direction is relatively small as compared to armchair direction. Apart from the anisotropic electronic conductance, BP$_5$ also presents a large value of the carrier mobility. Especially, the electron mobility at room temperature is up to 7.07×10$^3$ cm$^2$/V s, which is comparable or even higher that value of black phosphporene (2.30×10$^3$ cm$^2$/V s),[47,51,52] and significantly larger than other 2D materials, such as MoS$_2$,[53] making BP$_5$ very promising for application in electronic devices.

## Discussion

**Potential application.** So far, we have demonstrated that BP$_5$ monolayer is an intrinsic 2D multiferroic material with coupled ferroelasticity and ferroelectricity. Therefore, a potential application to electronic devices, particularly memory devices, will be possible by using such fascinating properties. Taking ferroelectric memory as an example, the initial states of spontaneous polarization direction is shown in the left panel of **Fig. 6**, which can be labeled as information '000'. By applying large electric field, one can switch the polarization direction and then the information '100' will be saved into the device, which is stored in a permanently accessible, nonvolatile form as the polarization state cannot be changed spontaneously. Considering its strongly anisotropic visible-light absorption and highly anisotropic carrier mobility of BP$_5$, the 'reading' process can be realized by either of three methods: 1) measuring current-voltage curve under small bias; 2) detecting photocurrent upon photo-illumination since the internal electric field will determine the drifting direction of photoexcited charge carriers; 3) measuring polarization-dependent photoluminescence. For ferroelastic memory, 'writing' process can be realized by external tensile stress and 'reading' by a similar way as in ferroelectric memory.

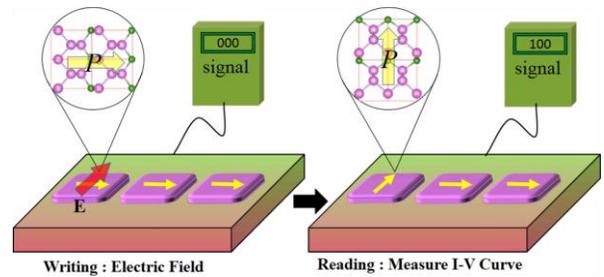

**Figure 6 | Potential application of BP$_5$.** Schematic of ferroelectric-induced rotation of spontaneous polarization direction in BP$_5$ when employed in nonvolatile memory device. The red arrow denote the direction of external electric fields, while the yellow arrow display the spontaneous polarization direction of BP$_5$ monolayer. The information '100' is saved into the device.

**Possible epitaxy substrate.** Although 2D BP$_5$ shows intriguing electro-mechanical properties for potential applications to nonvolatile memory, how to synthesize this material is still a critical issue. Therefore, we turn to investigate possible substrate for synthesizing BP$_5$ monolayer. AlN is reported to be an excellent substrate for growth of cubic boron phosphide. However, unlike the experimental one[20] where AlN (001) surface is used and cubic boron phosphide [111] orientation is predominant, we here find AlN (010) surface is a suitable substrate for epitaxy growth of BP$_5$. The optimized atomic configuration of AlN (010) surface with adsorbed BP$_5$ [marked as BP5@AlN(010)] is shown in **Fig. S7**. It can be found that BP$_5$ can sustain on the AlN substrate without notable distortion (mismatch < 5%). To guide future experiments, we also provide the STM image for BP5@AlN(010) (see **Fig. S7**).

In summary, based on global optimization Cuttlefish algorithm, we propose a new fascinating 2D material composed of boron and phosphorus element, namely BP$_5$. We find that BP$_5$ is a multiferroic material with coupled ferroelectricity and ferroelasticity, which could be used in nonvolatile memory devices in the future. Particularly, BP$_5$ possesses the highest reversible ferroelastic strain known till now and a ferroelectric polarization which is absent for black phosphorene. Moreover, BP$_5$ is an indirect semiconductor with a band gap of 1.34 eV and displays outstanding mechanical, optical and electronic properties, such as negative Poisson's ratio, strongly anisotropic visible-light absorption and high carrier mobility. Finally, AlN (010) surface has been demonstrated to be a potential substrate for epitaxy growth of BP$_5$. We hope our research can stimulate future experiments on this subject.

## Methods

The structure search is employed by global optimization Cuttlefish algorithm[24] (See Supplementary Information) for different stoichiometric $B_mP_n$ {m≤5 and n≤6 } compounds. Furthermore, we have reconfirmed the obtained structures of global minimum formation energy based on particle swarm optimization, as implemented in CALYPSO code.[54] All the first-principles calculations are performed on the basis of the Kohn-Sham density functional theory[55] (KS-DFT) as implemented in the Vienna ab initio simulation package[56] (VASP). The generalized gradient approximation as parameterized by Perdew, Burke and Ernzerhof[57] (PBE) for exchange-correlation functional is used to relax the geometric structures and the plane wave basis sets with kinetic energy cutoff of 450 eV are used to expand the valence electron wave functions. The convergence criterion for the energy in electronic SCF iterations and the force in ionic step iterations are set to be $1.0\times10^{-6}$ eV and $5.0\times10^{-3}$ eV/Å, respectively. A large vacuum space of at least 15Å is introduced to prevent interlayer interactions between periodic images. The reciprocal space is sampled with a k-grid density of 9×12×1 using the Monkhorst-Pack scheme. Ferroelectric spontaneous polarization is calculated by using the Berry phase approaches.[58,59] Besides, quasiparticle band structure is also calculated by many-body perturbation theory calculations with the $G_0W_0$ approximation[41,42] and optical properties are obtained by solving two-particle Bethe-Salpeter equation (BSE).[43,44] Van der Waals (vdW) correction proposed by Grimme (DFT-D2) is used due to its good description of long-range vdW interactions for multi-layered 2D materials.[60–64] The mechanical properties of materials, such as elastic constants, Young's modulus and Poisson's ratio[65–67] are calculated by our PyGEC package[9] with VASP interface.

## Acknowledgements

This paper is financially supported by the National Key Research & Development Program of China (Grant No. 2016YFA0200604), by the National Natural Science Foundation of China (NSFC) (Grants No. 21421063, No. 21233007, No. 21688102 and No. 21603205), by the Chinese Academy of Sciences (CAS) (Grant No. XDB01020300), by the Fundamental Research Funds for the Central Universities (Grant No. WK2060030023), by the China Postdoctoral Science Foundation (Grant No. BH2060000033). We used computational resources of Super-computing Center of University of Science and Technology of China, Supercomputing Center of Chinese Academy of Sciences, Tianjin and Shanghai Supercomputer Centers.

## Author contributions

H.W. and Y.J. conceived and demonstrated the initial idea of this research. H.W. J.S and Z.L. collected all the data. H.W., X.L. and Y.J. wrote the paper and all authors commented on it.

## Additional information

**Supplementary Information:** The detail of structure search Cuttlefish algorithm; energy evolution during the structure search; metastable allotropes of $BP_5$; snapshots of AIMD simulations; elastic constants, Young's modulus and Poisson's ratio; band structure and DOS under PBE level; equation for evaluate carrier mobility; $BP_5$ on AlN(010) surface and its STM image are provided.

**Competing financial interests:** The authors declare no competing financial interest.

# Supplementary Information for "BP$_5$ Monolayer with Multiferroicity and Negative Poisson's Ratio: A Prediction by Global Optimization Method"

Haidi Wang,[†] Xingxing Li,[†,‡] Jiuyu Sun,[†] Zhao Liu[†] and Jinlong Yang[†,‡,*]

[†]Hefei National Laboratory for Physical Sciences at the Microscale, University of Science and Technology of China, Hefei, Anhui 230026, China

[‡]Synergetic Innovation Center of Quantum Information & Quantum Physics, University of Science and Technology of China, Hefei, Anhui 230026, China

E-mail: jlyang@ustc.edu.cn

Supplementary Notes 1–3.

Supplementary Figures S1–S7.

Supplementary Tables S1.

Supplementary references.

## Supplementary Notes

### Supplementary Note 1: Cuttlefish algorithm

The structure search is employed by Cuttlefish algorithm,[1] which is a new meta-heuristic optimization algorithm that is inspired by the mechanism of color changing behavior of the cuttlefish to find the optimal solution in numerical optimization problems. The algorithm considers two main processes: *Reflection* and *Visibility*. Reflection process simulates the light reflection mechanism through the combination of three cell layers (Chromatophores, Iridophores and Leucophores),[2] while visibility simulates the visibility of matching patterns. These two processes are used as a search strategy to find the global optimal solution. The formulation of finding the new solution (*newP*) by using *reflection* and *visibility* is as follows:

$$newP = Reflection + Visibility$$

During the evolution, the population (also called cells) are divided into four groups ($G_1$, $G_2$, $G_3$ and $G_4$). For $G_1$, the algorithm applying case 1 and 2 (the interaction between chromatophores and iridophores) to produce a new solutions. These two cases are used as a global search. For $G_2$, the algorithm uses case 3 (Iridophores reflection operator) and case 4 (the interaction between Iridophores and chromatophores) to produces a new solutions as a local search. While for $G_3$ the interaction between the leucophores and chromatophores (case 5) is used to produce solutions around the best solution (local search). Finally for $G_4$, case 6 (reflection operator of leucophores) is used as a global search by reflecting any incoming light as it without any modification (Fig. S1). The main step of Cuttlefish algorithm is described as follows:

1 Initialize population (P[N]) with random solutions and assign the values of $r_1$, $r_2$, $v_1$, $v_2$.

2 Evaluate the population fitness and keep the best solution.

3 Divide population **into** four groups ($G_1$, $G_2$, $G_3$ and $G_4$).

4 Repeat

    4.1 Calculate the average **value** of the best solution.

    4.2 **for** (each element **in** $G_1$)

        generate **new** solution **using** Case(1 and 2)

    4.3 **for** (each element **in** $G_2$)

        generate **new** solution **using** Case(3 and 4)

    4.4 **for** (each element **in** $G_3$)

        generate **new** solution **using** Case(5)

    4.5 **for** (each element **in** $G_4$)

        generate **new** solution **using** Case(6)

    4.6 Evaluate the **new** solutions

5. Until (stopping criterion **is** met)

6. Return the best solution

Equations that are used to calculate reflection and visibility for the four groups are described as follow

$$R = random() * (r_1 - r_2) + r_2$$
$$V = random() * (v_1 - v_2) + v_2$$

where, random() is a function used to produce a random numbers between (0, 1). $r_1$, $r_2$ are two constant values used to find the stretch interval of the chromatophore cells. While $v_1$ and $v_2$ are two constant values used to find the interval of the visibility's degree of the final view of the pattern. The constants $r_1$, $r_2$, $v_1$ and $v_2$ in our calculation are set to be -2, 2, -2 and 2, respectively.

Case 1 and 2 for $G_1$:

$$Reflection[j] = R * G_1[i].Points[j]$$

$$Visibility[j] = V * (Best.Points[j] - G_1[i].Points[j])$$

Case 3 and 4 for $G_2$:

$$Reflection[j] = R * Best.Points[j]$$

$$Visibility[j] = V * (Best.Points[j] - G_2[i].Points[j])$$

Case 5 for $G_3$:

$$Reflection[j] = R * Best.Points[j]$$

$$Visibility[j] = V * (Best.Points[j] - AV_{best})$$

Case 6 for $G_4$:

$$P[i].Ponts[j] = Random * (upperLimit - lowerLimit) + LowerLimit$$

where $i$ presents the $i$-th element in group G, $j$ is the $j$-th point of $i$-th element in group G; *Best* is the best solution and *AV_best* presents the average value of the *Best* points; *upperLimit* and *lowerLimit* are the upper limit and the lower limit of the problem domain.

## Supplementary Note 2: Basic mechanical properties

We employ PyGEC[3] to calculate the mechanical properties of BP$_5$. The calculated four elastic stiffness constants $C_{11}$, $C_{12}$, $C_{22}$ and $C_{66}$ are listed in Table S1. The BP$_5$ displays anisotropic mechanical properties, since the shape of orientation-dependent Young's modulus (in-plane) is not a standard circle [See Fig. S5 (b)]. In details, the Young's modulus of the zigzag direction has a minimum value of 106.5GPa, while the maximum value (173.9 GPa) is along the armchair direction, which is smaller than those of graphene (1.0 *TPa*), MoS$_2$ (0.33*TPa*) and BN (0.25 *TPa*),[4–7] suggesting that 2D δ-P is relatively flexible. As shown in Figure S5 (c), the in-plane Poisson's ratio displays a normal ratio with maximum (minimum) value equals 0.22 (0.14), which means that BP$_5$ will contract in the other two lateral directions, when it is stretched along the armchair or zigzag direction.

## Supplementary Note 3: Deformation potential theory

The definition of carrier mobility for 2D systems can be written as:[8–11]

$$\mu_{2D} = \frac{eh^3 E}{(2\pi)^3 k_B T m_e^* m_d (E_l^i)^2}$$

where $e$ is the electron charge; h is the plank's constant; T is the temperature; $m_e^*$ is the carrier effective mass along the transport direction and $m_d$ is determined by $m_d = \sqrt{m_x^* m_y^*}$. The deformation potential constant of the VBM for hole along the zigzag direction defines as $E_l^{zig} = \Delta E/(\Delta l_x/l_{x,0})$, where $\Delta E$ is the energy change of VBM under the lattice compression and stretch from the equilibrium distance $l_{zig,0}$ by a distance of $\Delta l_{zig}$. The term $E$ is the elastic modulus of zigzag or armchair direction, which can be directly calculated by PyGEC. For armchair direction and CBM, $\mu_{2D}$ can be obtained similarly.

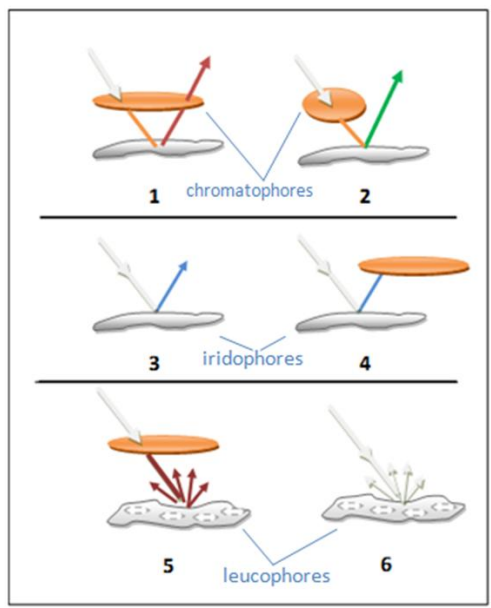

**Supplementary Fig. S1.** The six cases mimics the work of the three cell layers that are used by cuttlefish to change its skin colors.[1]

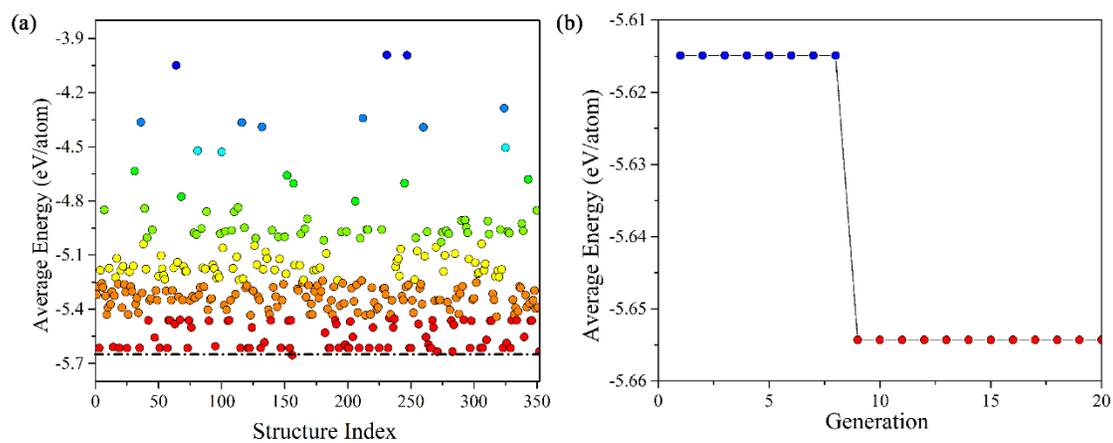

**Supplementary Fig. S2.** (a) All structures of BP$_5$ during the global search. (b) The energy evolution of BP$_5$.

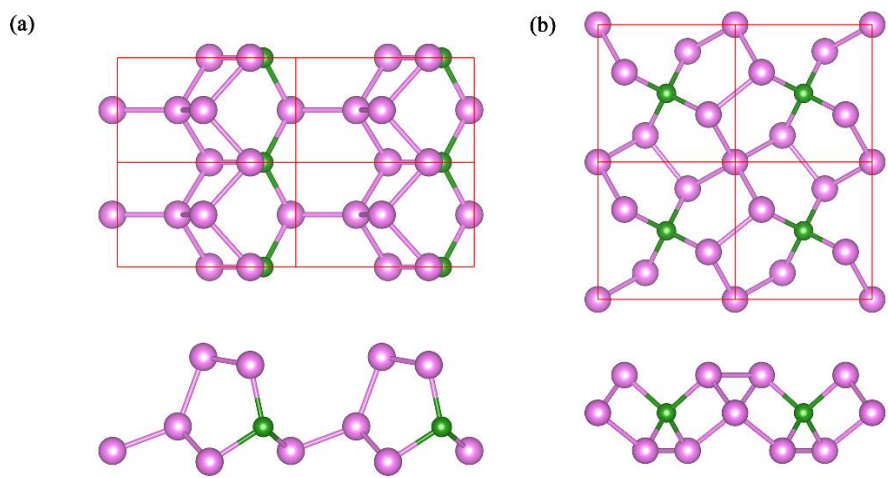

**Supplementary Fig. S3.** The top and side view of metastable allotropes of BP$_5$ with negative formation energy, β-BP$_5$ (a) and γ-BP$_5$ (b).

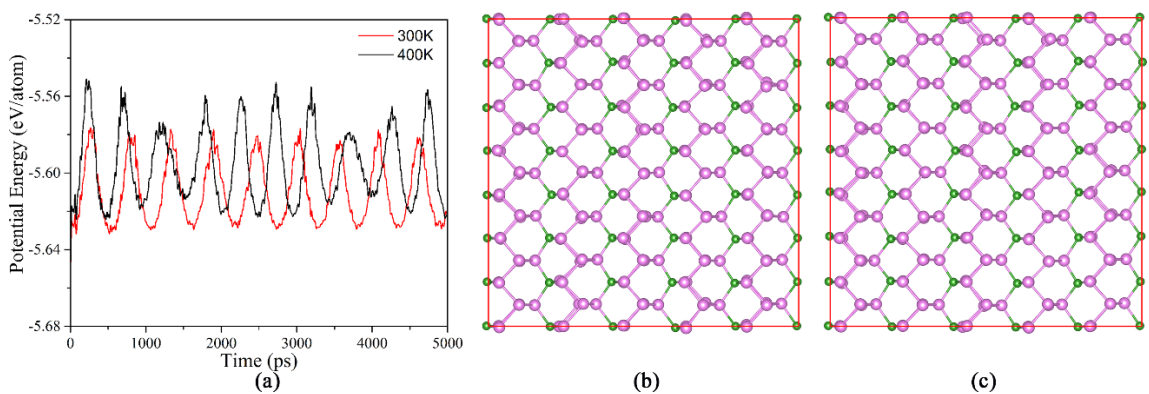

**Supplementary Fig. S4.** (a) Fluctuation of potential energy of BP$_5$ (5×7 supercell) during a NVT AIMD simulation at 300 K and 400K. (b) and (c) are snap of structure for BP$_5$ at the end AIMD simulation.

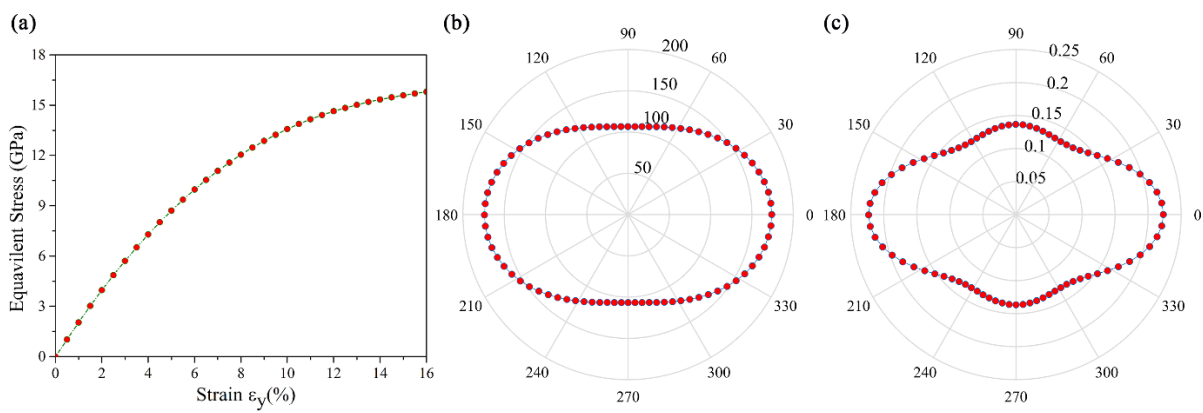

**Supplementary Fig. S5.** (a) Stress-strain relation for BP$_5$ while the stress is applied along the y (zigzag) direction. In-plane orientation-dependent (b) Young's modulus and (c) Poisson's ratio.

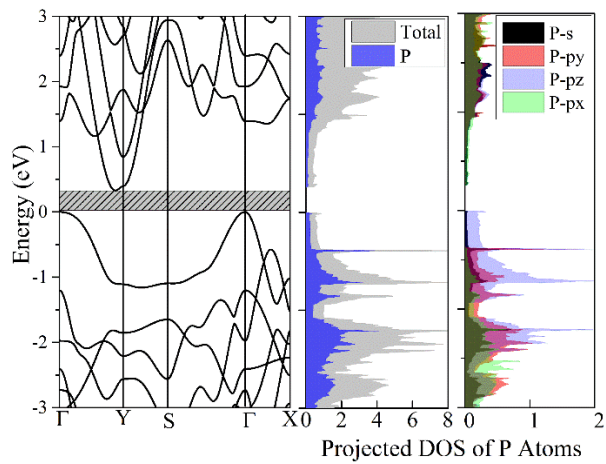

**Supplementary Fig. S6.** The band structure and projected density of states (PDOS) of BP$_5$ calculated by PBE functional.

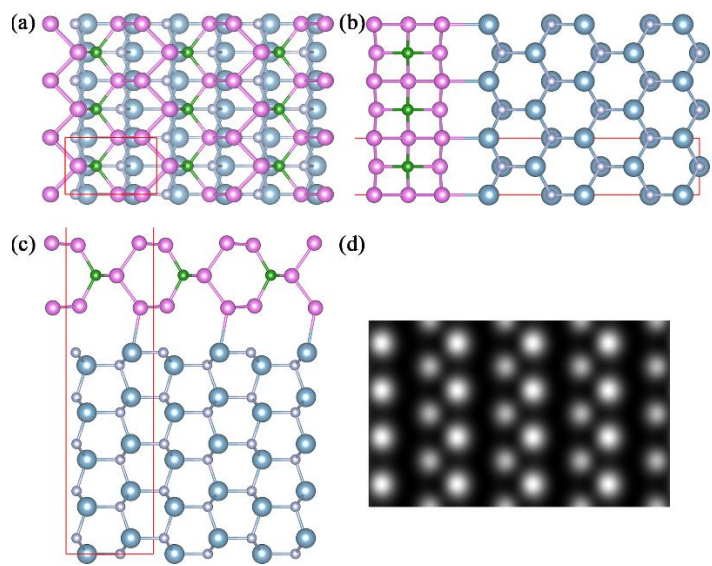

**Supplementary Fig. S7.** (a) (b) and (c) geometric structure of BP$_5$@AlN(010) surface, top and side view. (d) STM image of BP$_5$@AlN(010) with a bias voltage equals 2.5 eV.

**Supplementary Table S1.** The calculated elastic constants, Young's modulus along $x$ ($E_x$, armchair) and $y$ ($E_y$, zigzag) direction and Poisson's ratio along $x$ ($v_{xy}$) and $y$ ($v_{yx}$) direction of BP$_5$.

| Elastic constants /GPa | | | | Young's Modulus /GPa | | Poisson's ratio | |
|---|---|---|---|---|---|---|---|
| $C_{11}$ | $C_{12}$ | $C_{22}$ | $C_{66}$ | $E_x$ | $E_y$ | $v_{xy}$ | $v_{yx}$ |
| 179.37 | 24.52 | 109.88 | 60.32 | 173.90 | 106.53 | 0.22 | 0.14 |

**Supplementary references**